# Theoretical and practical challenges of using three ammeter or tree voltmeter methods in teaching

Vladimir Šimović, Trpimir Alajbeg, Josip Ćurković
Department of Electrical Engineering
Zagreb University of Applied Sciences
vsimovic@tvz.hr, trpimir.alajbeg@tvz.hr, jcurkovic@tvz.hr

**Abstract** - The tree ammeter method and the three voltmeter method are used for measurements of power. ore specifically, they are used to calculate the power factor of a specific load. oth methods are used on the „Fundamentals of electrical engineering" course in professional study of electrical engineering at the agreb niversity of Applied ciences as an introduction to measurements of power and phasor arithmetic. oth methods are susceptible to accuracy problems caused by small errors in measuring devices. These accuracy problems in specific scenarios that were used during the course presented an opportunity to educate students on the difference of theory and its practical application. This paper e amines our solution to the perceived accuracy problems and discusses the changes that were made in course material and teaching methods to highlight the difference between theory and practice.

## I. INTRODUCTION

As an integral part of the Fundamentals of electrical engineering, course students must attend and earn a passing grade in five laboratory exercises. The exercises are used as a tool to educate students on the process of conducting experiments, interpretation and discussion of results and applying theoretical knowledge from course classes. For many of the students, the laboratory exercises present the first real opportunity to handle instruments specific to electrical engineering.

Among the five laboratory exercises, one deals in measurements of power, specifically with the measuring and calculating the power factor ($\cos \varphi$) of a specific load ($Z_t$) in a simple AC circuit. This is done by using two methods; three ammeter and the three voltmeter method.

## II. THE METHODS

### A. Three voltmeter method

The three voltmeter is used in an inductive circuit to measure the value of the power factor. As seen on Figure 1, one voltmeter is used to measure the voltage of the circuit ($U_i$), the second one measures the voltage on the non-inductive resistance ($U_R$) that is connected in the series with the load branch and the third voltmeter is used to measure the voltage of the load ($U_t$).

The power (P) consumed by the load can be determined as the product of the voltage and current of the load ($U_t, I_t$) and the cosine of the phase angle between these two (the power factor).

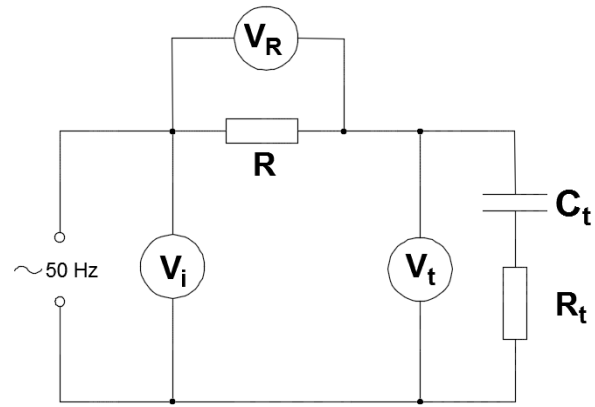

Figure 1.  The three voltmeter method

$$P = U_t \, I_t \cos \varphi \tag{1}$$

Using the law of cosines, the power factor ($\cos \varphi$) can be calculated.

$$\cos \varphi = (U_i^2 - U_R^2 - U_t^2) / (2 \, U_R \, U_t) \tag{2}$$

For optimal accuracy, the non-inductive resistance should be large enough so that the voltmeter (or multimeter) can measure it with satisfactory accuracy, but not too large, otherwise the voltage available to the load would be too small. Ideally, it should be close or equal to load impedance.

### B. Three ammeter method

The three ammeter is also used in an inductive circuit to measure the value of the power factor, independent of source frequency and waveforms. In this method, as seen on Figure 2, across the inductive circuit load in which the power factor is to be determined, a non-inductive resistance is connected parallel with the load branch. One ammeter is used to measure total current of the circuit, the second one measures current going through the non-inductive resistance and the third ammeter measures the current of the load branch.

Current through the resistive branch ($I_R$) is in phase with source voltage while the current of the load ($I_t$) has its own phase, which will affect the power factor (cos φ). Total current ($I_i$) is a vector sum of the other two currents. Using the law of cosines, the power factor can be calculated.

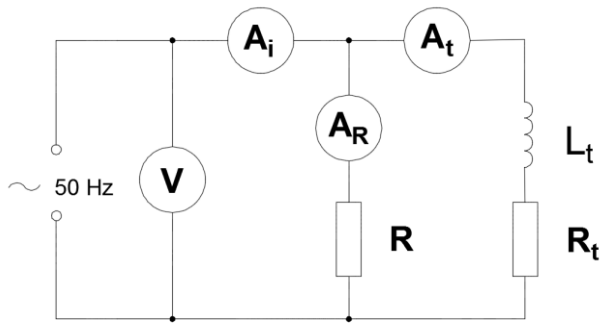

Figure 2. The three voltmeter method

$$\cos \varphi = (I_i^2 - I_R^2 - I_t^2) / (2\ I_R\ I_i) \tag{3}$$

For optimal accuracy, ideally, the non-inductive resistance should be close or equal to load impedance as it should be for the three voltmeter method.

## III. METHOD IMPLEMENTATION

### A. The actual configuration

In these experiments for measuring the power factor in both three ammeter and three voltmeter methods the AC source is set to 7,5 V RMS (Root-Mean-Square), with the frequency set at 50 Hz. The voltmeter that is connected in parallel to the source, as seen on Figures 1 and 2, is used to set its output voltage. For both methods, the same electrical circuit elements were used; a non-inductive resistance value of 680 Ω and a load phase that consisted of a variable resistor of 100 Ω and a capacitor of 10μF. Three Digital Multimeters PeakTech 2010 were used to measure current or voltage, depending on the method. The used elements were connected as seen on Figures 1 and 2. For the three ammeter method a fourth Digital Multimeters PeakTech 2010 was used to set the voltage of the source.

### B. Occurring problem

One of the problems that affected the results of both methods was the fact that in the original configuration the measurement devices that were used were all analog. This was mostly due to the fact that students were divided in six pairs that did the laboratory exercise simultaneously. This meant that at any given time during the class, at least eighteen instruments are needed. Since analog meters use a needle and scale to indicate values, the accuracy of the results is also affected by the operator's ability to read the readout on the meter, as seen on Figure 3.

There is also the problem of the meter's impedance. Digital voltmeters have significantly higher impedance than their analog counterparts; they are more accurate when measuring voltage.

It is well documented [7][8][9] that for optimal accuracy the methods require that the non-inductive resistance should be close or equal to load impedance.

In the original configuration of the laboratory exercise, an adjustable resistor (rheostat to be specific) was used in the two-part load impedance. As a wire-wound resistor, a rheostat has a large inductance at higher frequencies, which affects the impedance of the load and its power factor, which would translate into further deviation from the theoretical one. Since the used frequency is 50 Hz, the effect in negligible.

There is also the matter of internal impedance of circular wires that for most low-frequency applications has no effect on the circuit or the results of the experiment and therefore can be ignored.

### C. Presented opportunity

The noted problems presented an opportunity to further educate the students on the disparity between the practical and theoretical. To be more specific, to educate them about:

- risks of misreading analog instruments
- effects of measurement devices on the measurements themselves

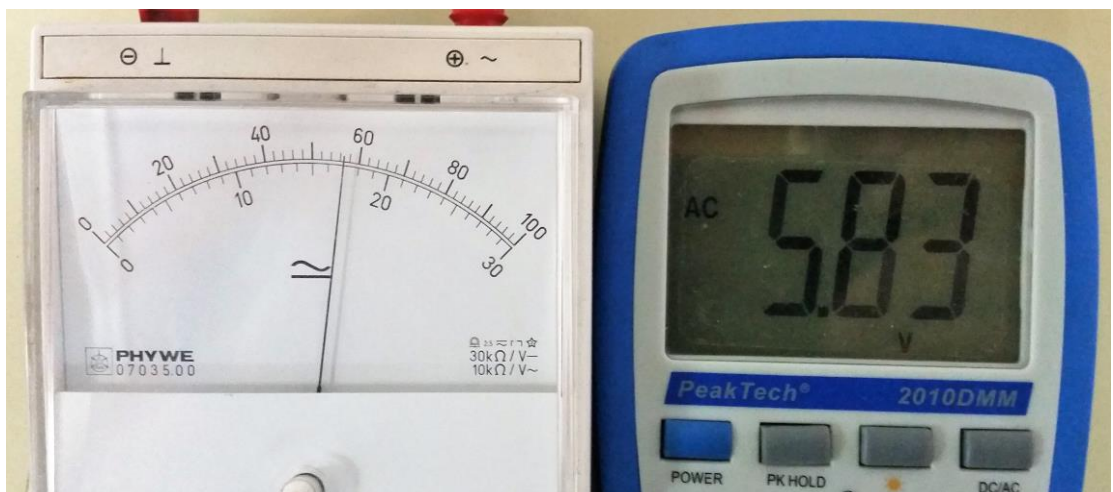

Figure 3. Example of measuring uncertainty for analog and digital measuring device

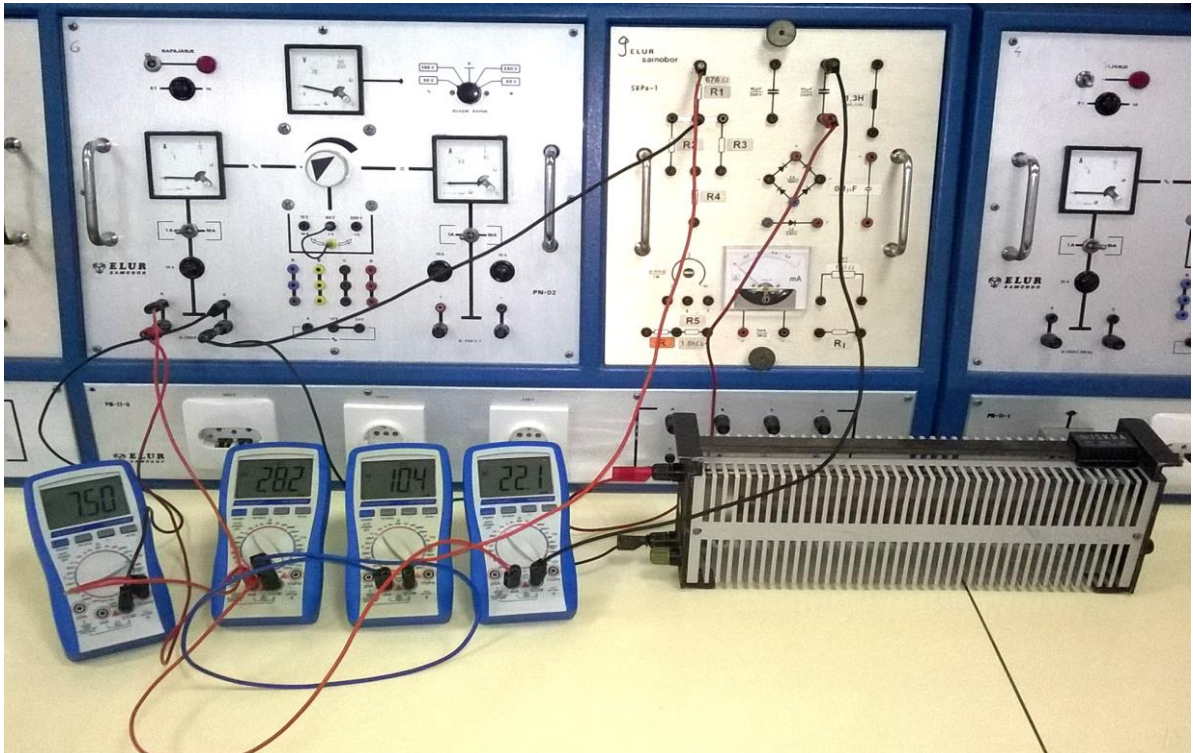

Figure 4. The actual configuration of the experiment with the three ammeter method used as described

- understanding instrument and circuit components specifications

- parasitic elements of electrical components

- effects of change in AC frequency

## IV. CHANGE IN TEACHING METHODS

In the process of tackling the occurring problems different angles had to be considered. In the end, changes were made in all stages of this specific laboratory exercise and even some other exercises.

### A. re e periment

As part of the preparation for the laboratory exercise, students are assigned problems to solve to better prepare themselves for the exercise itself. With effects of measurement devices on the measurements themselves in mind, a specific circuit, similar to the one in the experiments, was assigned. Students were instructed to calculate the deviation in measurement devices readouts between two cases; when the instruments were ideal (no internal impedance) and when they were real (internal impedance was given). The assignments were done for a specific frequency, but can be altered and expanded for calculation for a wide range of frequencies using Microsoft Excel or some other spreadsheet program (Open Office Calc comes to mind as an open source alternative). This approach requires some knowledge of using functions in a spreadsheet program, which makes it an ideal task for a different course called Personal Computer Applications that is also taught as a part of professional study of Electrical engineering at the Zagreb University of Applied Sciences.

### B. During the e periment

For the laboratory experiment itself, few key elements were changed.

Since uncertainty of reading analog instruments or human error is not a part of the intended lesson, nor the curriculum of Fundamentals of electrical engineering, analog instruments were swapped with their digital counterparts. These subjects are a part of a different course; Electrical Measurements, and therefore, because of the change, teaching materials have been altered for a different laboratory exercise on that course to educate the students on accuracy problems related to reading of analog instruments.

The most important change is the change in load impedance. Instead of the wire-wound resistor (rheostat), a typical carbon composition resistor was used. This lessened the effect of the rheostat's large inductance on the impedance of the load and its power factor but more importantly, because of its small size, the swap enabled us to construct eight different load phase impedances that were placed in 3D printed black boxes This was done to insure that all students will have different measurement results which among other things would reduce cheating. The changes are noted on Figures 5 and 6.

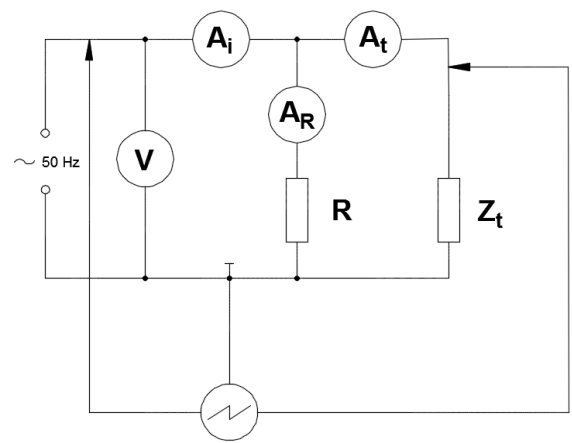

Figure 5. Updated experiment using the three ammeter method

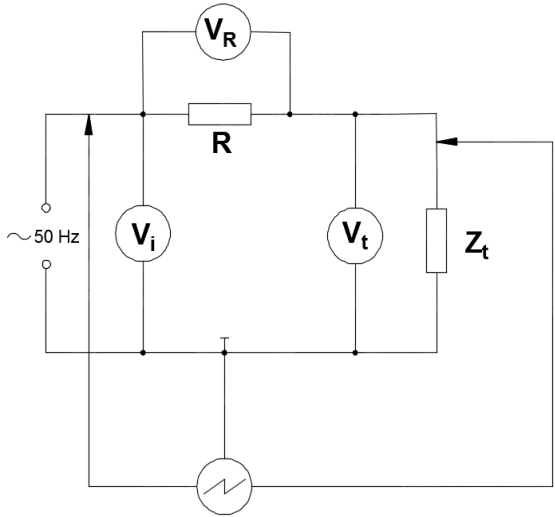

Figure 6. Updated experiment using the three voltmeter method

In addition, an oscilloscope was added to the experiment with one channel connected to the AC source and the other channel to the load phase. The oscilloscope is used for both methods but is crucial to the three voltmeter method where it is used to teach students about phase shifts. The students are instructed to compare the value calculated from the power factor and the value of the phase shift that can be read from the oscilloscope. By doing so, they can determine the nature of the load, more specifically weather it is inductive or capacitive in nature. With the usage of a digital oscilloscope, the readout can be stored for analysis on the computer at home.

### C. ost e periment

As a tool for further education of the students on all of the above-mentioned topics, an EDA (Electronic Design Automation) program is used post-experiment. The Proteus ISIS (Intelligent Schematic Input System) package is the EDA program of choice. Students can freely and legally download a demonstration version of the program. Among several other restrictions, the main restriction, in comparison to a full version, is that the project cannot be saved or stored digitally in any way, but the readouts can be manually noted in the post-experiment documentation. These can be used later for comparison with the results of the pre-experiment assignment and the experiment itself.

The Proteus ISIS package, as seen on Figure 7, enables further education post-experiment in a way that could not be realised during the laboratory exercise because of time constraints. With Proteus, the student can simulate the behaviour of the circuit and its elements on a wider frequency spectre without the need for additional physical circuit elements. To execute that in a laboratory exercise environment, an addition of a signal generator would be needed and perhaps even a data acquisition module. This would further complicate the exercise itself for both the student and the educator.

Changing the values of electric circuit elements is an easy task in Proteus and that helps to evaluate both methods in sub-optimal configurations, specifically, when the non-inductive resistance is not close or equal to load impedance. In addition, changing the load impedance in Proteus, as with any other element in the circuit, results in an instant change of instrument readouts. This is something that yet again, is not so easy to accomplish in the laboratory environment. Not only because of time constrictions, but because it is not practical to have so many circuit elements at student's disposal. So, instead of having multiple sets of capacitors, inductors and resistors, physically available, to evaluate the behaviour of the method when the non-inductive resistance is not close or equal to load impedance, this can be achieved by simply adjusting the elements in Proteus with a few clicks of the mouse.

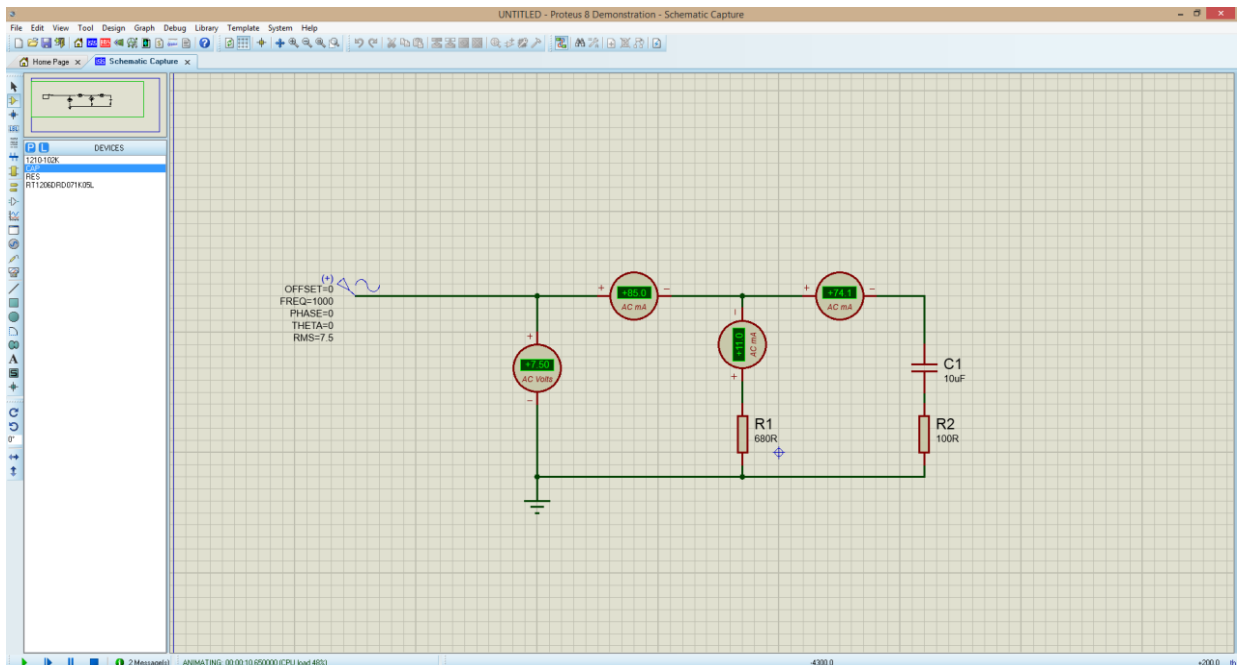

Figure 7. Proteus ISIS interface showing a simulation of the three ammeter experiment

## V. CONCLUSION

Although power factor meters exist, even today, the tree ammeter method and the three voltmeter method are commonly used for calculation of the power factor of a specific load, both in practice and as a teaching tool. Both methods have practical and theoretical downsides and limitations, those same downsides and limitations can be used as a part of the teaching process in educating students on the divergence of the practical from the theoretical.

The addition of an EDA program proved to be an invaluable asset and a teaching tool even for a „Fundamentals of electrical engineering" course in professional study of electrical engineering at the Zagreb University of Applied Sciences, where up until recently, the usage of computers, for this specific course, was a rarity. Given the methods susceptibility to accuracy problems outside specific conditions described in section II, and physical and time constraints of working with students in a laboratory setting, the addition of an EDA program after the exercise helped us immensely to further educate the students about the problems when the mentioned methods were applied in sub ideal conditions.

Given the satisfaction with the involvement of computer usage in the education process in the mention course, further developments and EDA program inclusions are planned.